\documentstyle[aps,pre,twocolumn,epsf,graphicx,amsmath,array]{revtex}
\begin{document}
\newcommand{\btau}{\mbox{\boldmath{$\tau$}}}
\twocolumn[\hsize\textwidth\columnwidth\hsize\csname@twocolumnfalse%
\endcsname
\title{Geometry of frictionless and frictional sphere packings}
 
\author{Leonardo E.~Silbert$^1$, Deniz Erta{\c s}$^2$, Gary
  S.~Grest$^1$, Thomas C.~Halsey$^2$, and Dov Levine$^3$}
 
\address{$^1$ Sandia National Laboratories, Albuquerque, New Mexico 87185\\ 
  $^2$ Corporate Strategic Research, ExxonMobil Research and
  Engineering, Annandale, New Jersey 08801\\ $^3$ Department of
  Physics, Technion, Haifa, 32000 Israel}

\date{\today}
 
\maketitle                                                                      
\begin{abstract}
  We study static packings of frictionless and frictional spheres in
  three dimensions, obtained via molecular dynamics simulations, in
  which we vary particle hardness, friction coefficient, and
  coefficient of restitution. Although frictionless packings of
  hard-spheres are always isostatic (with six contacts)
  regardless of construction history and restitution coefficient,
  frictional packings achieve a multitude of hyperstatic packings that
  depend on system parameters and construction history. Instead of
  immediately dropping to four, the coordination number reduces
  smoothly from $z=6$ as the friction coefficient $\mu$ between two
  particles is increased.
\end{abstract}


]

\section{Introduction}

Dense amorphous packings of frictionless spheres have proven to be an
extremely useful paradigm in different physical contexts, such as
metallic glasses\cite{angell1}, colloidal crystals\cite{weitz1}, and
emulsion rheology\cite{lacasse1}. Granular materials are another
example of a system with macroscopically large particles, with one
major difference: grain-grain interactions involve frictional forces.
As a result, granular packings may be quite different from
frictionless sphere packings in ways which may impact significantly on
their physical properties.

A common quantity of interest in packings of hard spheres is the
average number of contacts per particle (coordination number) $z$.
In order to achieve static mechanical equilibrium,
each sphere in the packing needs a sufficient number of constraints
that freeze out its translational and rotational degrees of freedom.
These constraints are provided by contacts, and once there are
a sufficient number of them, the packing can accommodate external body or
boundary forces, as long as a set of contact forces satisfying
mechanical equilibrium can be found for the given arrangement of such
contacts. The minimal average coordination number required to obtain
static packings of $d$-dimensional frictionless spheres that are
stable against external perturbations is $z_{n}=2d$ \cite{alexander1},
whereas for spheres with friction, $z_{f}=d+1$ \cite{edwards2}. In
three dimensions, $z_{n}=6$ and $z_{f}=4$. We call such packings
``isostatic''.

In this study, we investigate whether or not sphere
packings readily achieve isostaticity under generic packing
conditions. This isostaticity hypothesis is important in
theories focusing on the macroscopic response of
such packings\cite{grinev1},and at first appears
reasonable, given the strong numerical evidence from simulation
\cite{lacasse1,makse1} that $z_{n}=6$ for dense random
packings of frictionless spheres. Simulation studies of frictional
spheres compressed in a gravity-free environment\cite{makse1} have
shown that $z_{f}$ is significantly less than 6, but with the 
lowest achieved value of around 4.5, it remains unclear whether 
the minimal value of 4 is reached in the limit of zero confining 
pressure (equivalent to the hard-sphere limit.) It is also
unclear in what way the packings would change for arbitrarily small 
friction coefficient $\mu$ between the spheres in order to achieve
a reduction in $z$ to four, if the isostaticity hypothesis were true.

To address such questions, we perform a systematic simulation study of
the effect of various parameters on sphere packings. In particular, we
vary the following materials properties: the sphere hardness $k_{n}$, 
the coefficient of restitution $\epsilon$, and the friction coefficient 
$\mu$ between two particles. We also vary the initial conditions of the 
packing by varying the initial packing density $\phi^i$, as well as
the initial velocities of the spheres. We investigate how the density, 
coordination number and the nature of
the contacts change as these parameters are varied. Although
frictionless hard-spheres appear to form isostatic
packings regardless of construction history and restitution
coefficient, frictional hard-spheres achieve a multitude of hyperstatic
packings $(z>z_f)$ that depend on system parameters and construction history
\cite{leo10}. The coordination number reduces smoothly from $z=6$ as
the friction coefficient is increased, disagreeing with the
isostaticity hypothesis.

\section{Simulation Method}

\begin{figure}[!]
\begin{center}
\includegraphics[width=2.7cm]{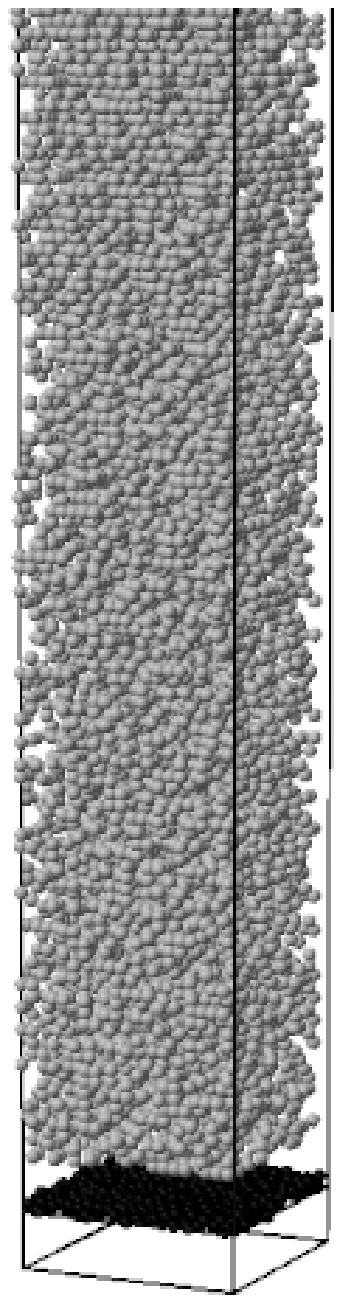}
\includegraphics[width=2.7cm]{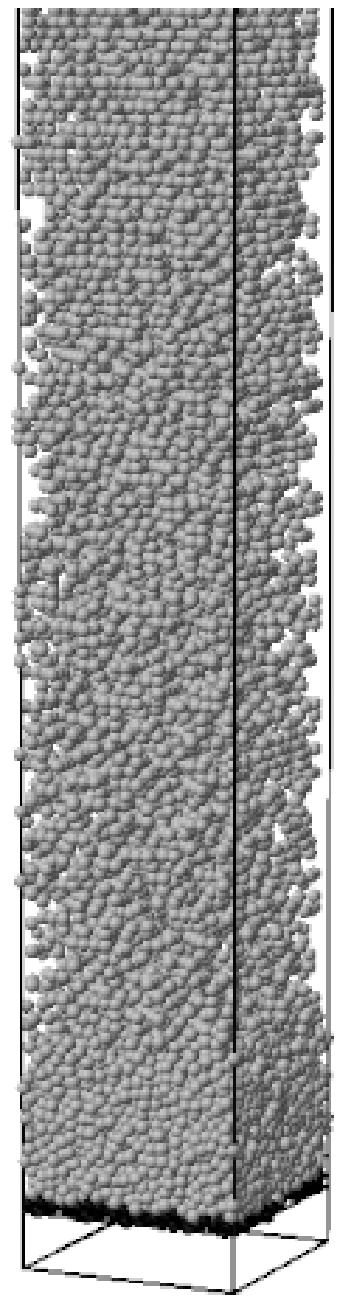}
\includegraphics[width=2.7cm]{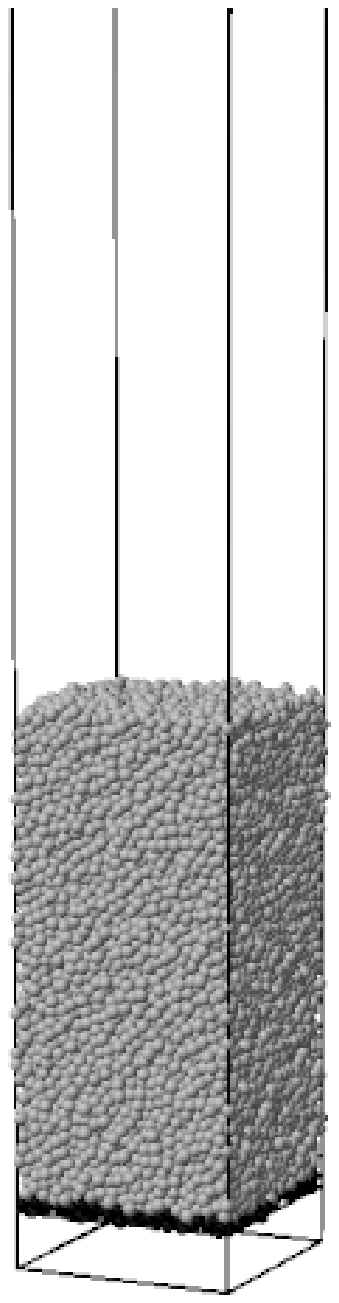}\\
(a)\hfil\hfil\hfil(b)\hfil\hfil\hfil(c)\\
\bigskip
\caption{Lower portion of the packing of $N=20,000$ spheres in a periodic cell
  $20{\textrm d}\times20{\textrm d}$ supported by a rough bed (black
  particles), constructed by settling under gravity, with $\mu=0.50$, 
  $k_{n}=2\cdot10^{5}mg/d$, and $\epsilon=0.88$. (a) Initial
  configuration with volume fraction $\phi^{i}\approx0.13$.(b)
  Intermediate time during settling. (c) Final (static) configuration
  with $\phi^{f}\approx0.60$. The black frame is added as a guide to the eye.}
\label{figure1}
\end{center}
\end{figure}

We present molecular dynamics (MD) simulations in three dimensions on 
model systems of $N=20,000$ monodisperse, cohesionless spheres of diameter 
$d$ and mass $m$. The system is spatially periodic in the $xy-$plane, with a
unit cell of size $20d\times 20d$, and is bounded in the $z-$direction
by a rough bed at the bottom and an open top. The starting
configurations consist of randomly positioned non-overlapping spheres, 
with packing fractions in the range $0.02<\phi^{i}<0.3$, obtained by 
varying the overall height while keeping $N$ and the $x,y$ dimensions 
fixed. The system is subsequently allowed to settle under gravity
on top of the rough bed (see Fig.~\ref{figure1}). The equilibrated
static packing height is about $50d$. This method of construction
mimics the pouring of granular materials through a sieve to an area
far away from side walls, without forming a conical heap.

Different ways of preparing static granular packings include
compressing them\cite{makse1}, and reducing the
inclination angle of gravity-driven chute flows below the angle of
repose\cite{leo10}. In the frictionless case, conjugate gradient
methods have been used to study dense random packings
\cite{lacasse1,ohern1}.  However, for the case of particles with
friction, MD simulations are more appropriate in order to properly
account for both the normal and tangential forces, since the latter
depend on the loading history of the contact.

The spheres interact only on contact through a linear spring--dashpot
interaction law \cite{footnote7} in the normal and tangential
directions to their lines of centers \cite{walton1,cundall1}.
Contacting spheres $i$ and $j$ positioned at ${\bf r}_{i}$ and ${\bf
  r}_{j}$ experience a relative normal compression $\delta=|{\bf
  r}_{ij}-d|$, where ${\bf r}_{ij}={\bf r}_{i}-{\bf r}_{j}$, which
results in a force
\begin{equation}
{\bf F}_{ij}={\bf F}_{n}+{\bf F}_{t}.
\label{equation1}
\end{equation}
The normal and tangential contact forces are given by
\begin{equation}                                             
{\bf F}_{n}=k_{n}\delta{\bf n}_{ij}-\frac{m}{2}\gamma_{n}{\bf v}_{n},
\label{equation2}
\end{equation}
\begin{equation}
{\bf F}_{t}=-k_{t}\Delta{\bf s}_{t}-\frac{m}{2}\gamma_{t}{\bf v}_{t},
\label{equation3}
\end{equation}
where ${\bf n}_{ij}={\bf r}_{ij}/r_{ij}$, with $r_{ij}=|{\bf
  r}_{ij}|$, ${\bf v}_{n}$ and ${\bf v}_{t}$ are the normal and
tangential components of the relative surface velocity, and $k_{n,t}$
and $\gamma_{n,t}$ are elastic and viscoelastic constants
respectively. $\Delta{\bf s}_{t}$ is the elastic tangential
displacement between spheres, obtained by integrating surface relative
velocities during elastic deformation of the contact. The magnitude of
$\Delta{\bf s}_{t}$ is truncated as necessary to satisfy a local
Coulomb yield criterion $F_{t} \leq \mu F_{n}$, where $F_{t} \equiv
|{\bf F}_{t}|$ and $F_{n} \equiv |{\bf F}_{n}|$. Frictionless spheres
can be simulated simply by setting $\mu=0$. Finally, the particles are
moved according to the total forces and torques applied to them
through contacts and the gravitational field. Additional details can
be found in Ref.\cite{leo7}.

The presented simulations were carried out for a range of materials
parameters. The normal spring constant $k_{n}$ varied from
$2\cdot10^{5}$ to $2\cdot 10^{9} mg/d$ in order to understand how
the system behaves as it approaches the hard sphere limit. In all
cases, $k_{t}=2k_{n}/7$ \cite{wolf1}. In order to study the crossover
from frictionless to frictional systems, the local particle friction
coefficient $\mu$ was varied from $0$ to $10$. Finally the coefficient
of restitution $\epsilon$, i.e. the ratio of the final to initial
normal velocities in a head-on binary collision, was also varied. For
a linear spring--dashpot interaction,
\begin{equation}
\epsilon=\exp(-\gamma_nt_{\rm col}/2),
\end{equation}
where the collision time $t_{\rm {col}}$, 
\begin{equation}
t_{{\rm col}}=\pi(2k_{n}/m - \gamma_{n}^{2}/4)^{-1/2}.
\label{tcol}
\end{equation}
Three values of $\epsilon$=$0.26$, $0.50$, and $0.88$ were used.  The
effect of $\epsilon$ is like that of varying the quench rate in a
thermal system: for smaller values of $\epsilon$, collisions are more
inelastic, energy is dissipated faster, and the spheres have less
ability to move from the the first position where they are
mechanically stable, which may result in packings that are less stable
than similarly prepared packings of more elastic grains.

Most of the simulations were started from a static configuration of
$\phi^{i}\approx0.13$ as shown in Fig.~\ref{figure1}(a). The timestep
$\delta t\approx 0.05\sqrt{m/k_{n}}$ was chosen to accommodate the
decreasing collision time as the particle hardness $k_{n}$ is
increased [cf. Eq.~(\ref{tcol})]. For $k_{n}=2\cdot10^{5}mg/d$,
$\delta t \approx 10^{-4}\sqrt{d/g}$. Simulations were then run until
the kinetic energy per particle was less than $10^{-8}mgd$ for small
$k_{n}$, and up to three orders of magnitude less for large $k_{n}$.
This requires $3-8\cdot10^{6}\delta t$ for small $k_{n}$, and
$4-8\cdot10^{7}\delta t$ for $k_{n}=2\cdot10^{9}mg/d$. 

Because of the
large amount of time required to reach static equilibrium, each set of
parameters was run for 1 to 5 configurations. In some cases, the
particles were given an initial, random velocity with an average KE
per particle of approximately $20-100 mgd$. The results were identical
in all cases within sample to sample fluctuations.

\section{Results}

\subsection{Coordination Number}
\label{seccoord}

\begin{figure}[h]
\begin{center}
\includegraphics[width=7.5cm]{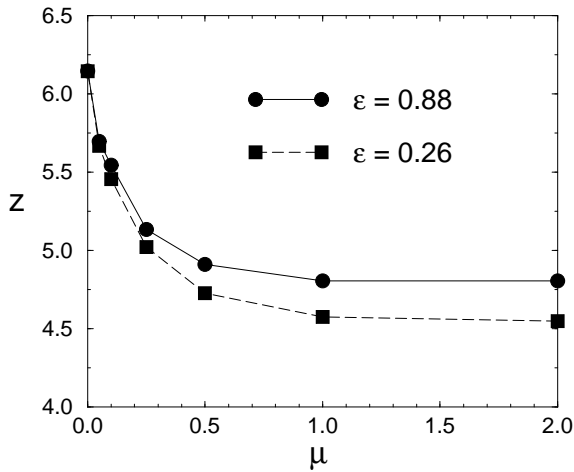}
\bigskip
\caption{Bulk averaged coordination number $z$ as a function of $\mu$ for
  $k_{n}=2\cdot 10^{5}mg/d$ and $\phi^{i}=0.13$ for two values of $\epsilon$.}
\label{zvsmu}
\end{center}
\end{figure}

\begin{figure}[h]
\begin{center}
\includegraphics[width=7.5cm]{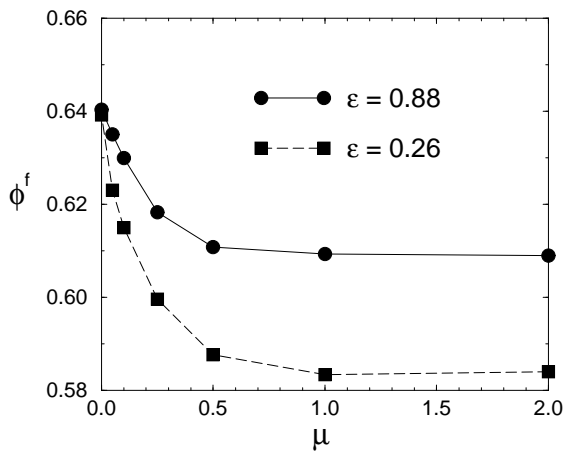}
\bigskip
\caption{Final volume fraction $\phi^f$ versus $\mu$ for the same parameters as in
  Fig.~\ref{zvsmu}.}
\label{phivsmu}
\end{center}
\end{figure}

Figure \ref{zvsmu} shows the effect of friction coefficient $\mu$ on
$z$ for $k_n=2\cdot 10^{5} mg/d$ and $\epsilon=0.88$ and $0.26$. For
both values of $\epsilon$, $z=6.144\pm0.002$ for frictionless packings. 
As we shall see in Sec.~\ref{sechard}, the deviation from the isostatic
value of 6 can be attributed to the finite stiffness of the spheres,
the isostatic value is apparently obtained in the hard-sphere limit.
However, there is no sudden drop from $z=6$ as friction is turned on;
rather there is a gradual decrease in $z$ to a parameter dependent
minimal value, accompanied by a similar decrease in the final volume
fraction $\phi^f$ (see Fig.~\ref{phivsmu}.) 

\vspace{-0.2in}

\begin{figure}[h]
\begin{center}
\includegraphics[width=7.5cm]{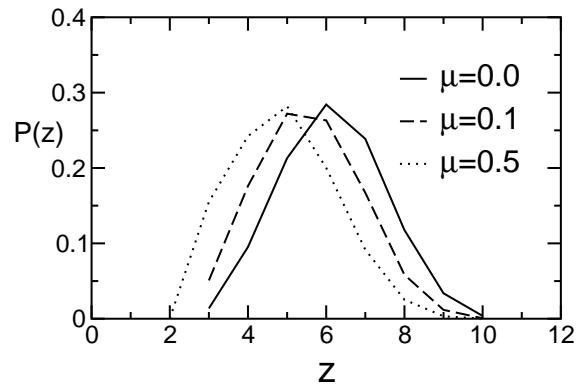}
\bigskip
\caption{Distributions of the coordination number shift to lower values
of $z$ as $\mu$ is increased. These results are for $k_{n}=2\cdot 10^{5}mg/d$,
 $\epsilon=0.88$ and $\phi^i=0.13$.
}
\label{zdist}
\end{center}
\end{figure}

As depicted in Fig.~\ref{zdist}, the decrease in $z$ is primarily due 
to an overall shift in the distribution of coordination numbers to 
lower values, rather than a change in its shape and width. 
Consequently, the frequency of particles with eight or more 
neighbours reduces as $\mu$ increases, and particles with as few 
as two contacts start to appear at $\mu=0.5$, 
indicative of arching within the packing at large $\mu$.
The saturation of $z$
and $\phi^f$ for $\mu\agt 1$ is due to the fact that the typical
tangential forces $F_t$ in a packing with $\mu=\infty$ is expected to
be of order $F_n$, and lowering $\mu$ has little effect on the
packing down to $\mu\approx 1$. This behavior of the contact forces
is further verified in Sec.~\ref{secplastic}.  

Unlike the frictionless case, the deviations from 
isostaticity $(z=4)$ for $\mu > 0$ cannot be attributed to corrections 
due to the finite stiffness of the spheres. The packings remain 
unambiguously hyperstatic in the hard-sphere limit.  

When a static packing with a surface tilt near the angle of repose was
generated by cessation of flow down an inclined plane\cite{leo10},
similar results for $z$ and $\phi$ were obtained. For
$k_{n}=2\cdot10^{5}mg/d,~\epsilon=0.88, ~{\rm and}~\mu=0.50$, such
packings gave $z = 4.69$ and $\phi=0.594$, compared to $z=4.90$ and
$\phi=0.61$ for packings presented in this study.

\subsection{The Radial Distribution Function}

\begin{figure}[h]
\begin{center}
  \includegraphics[width=7.5cm]{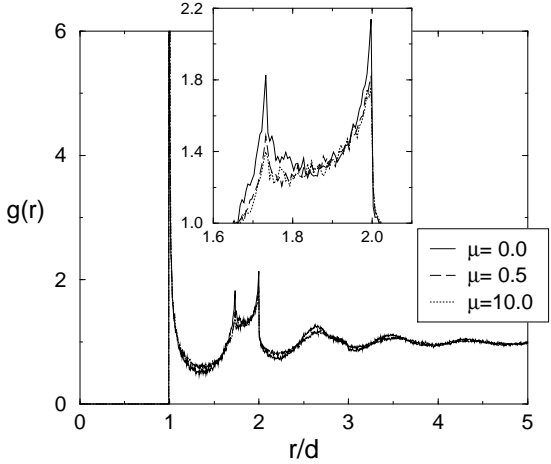} 
\caption{The radial distribution function $g(r)$ for spheres with and without
  friction for $k_{n} = 2\cdot10^{5}mg/d$ and $\epsilon = 0.88$.}
\label{gofrfric}
\end{center}
\end{figure}

The radial distribution function (RDF), $g(r)$, for
$k_{n}=2\cdot10^{5} mg/d$ and $\epsilon=0.88$, is plotted in
Fig.~\ref{gofrfric} for several values of $\mu$. The characteristic
split second peak, indicating short-range order out to second
neighbors, is evident. For $\mu=0$, $g(r)$ is essentially identical to
that obtained for random close packing (RCP), at volume
fraction $\phi^f\approx0.64$\cite{berryman}.  
As $\mu$ increases, the secondary peaks in $g(r)$
diminish, as seen in the inset. 

\begin{figure}[h]
\begin{center}
  \includegraphics[width=7.5cm]{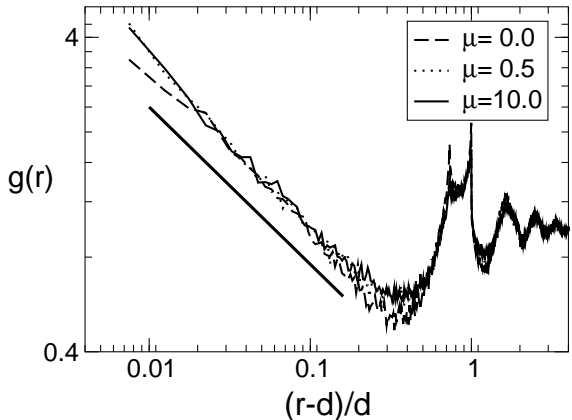} 
\caption{A logarithmic plot of $g(r)$ vs. $(r-d)/d$ (for $k_{n} = 2\cdot10^{5}mg/d$ 
and $\epsilon = 0.88$) reveals a power-law singularity with exponent 
$\alpha\approx0.5$ for both frictionless and frictional spheres.
Straight line has a slope of -1/2. The same power-law is observed up to
the largest values of $k_n$ studied.}
\label{grfirstpeak}
\end{center}
\end{figure}

The first peak of $g(r)$ is of particular interest, since near-contacts 
with $r/d$ just over 1 play an important role in the dependence of $z$ on 
the stiffness of the spheres (See Sec.~\ref{sechard}). Figure \ref{grfirstpeak} 
reveals a square-root singularity of the RDF near $r/d=1$, i.e.,
\begin{equation}
g(r) \propto \left(\frac{r}{d}-1\right)^{-\alpha}, \ 0 < \frac{r-d}{d} \ll 1, 
\label{eqgrlimit}
\end{equation}
with $\alpha=0.52\pm0.03$. This singularity has apparently not been reported 
elsewhere, although we have also verified its presence in the RDF of 
hard-sphere packings provided to us\cite{torquato}. Note that this singularity
is integrable, and is distinct from particles actually in contact.

\subsection{The Hard-Sphere Limit}

\label{sechard}

\begin{figure}[h]
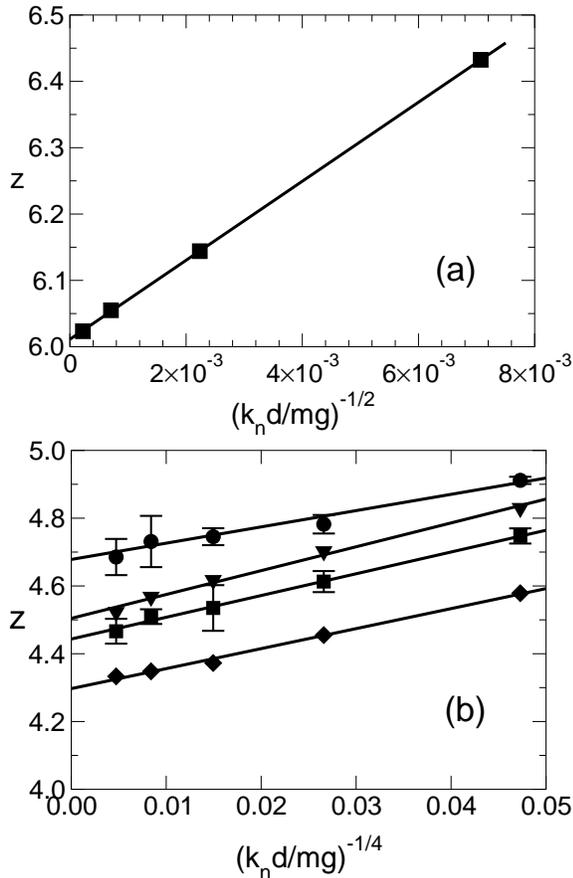

\begin{center}
\includegraphics[width=7.5cm]{zc_kn_nofric} \\
\includegraphics[width=7.5cm]{zc_kn_fric} 
\caption{Bulk averaged coordination number $z$ as a function of particle
  hardness $k_{n}$, for $\phi^{i}=0.13$: 
  (a) For frictionless spheres, where the 
  extrapolation to hard-spheres implies isostatic packing (z=6); 
  (b) For $[\mu,\epsilon]$ = [0.50, 0.88] (solid circles), [0.50, 0.50]
  (triangles), [0.50, 0.26] (solid squares), and [10.0, 0.26] (solid
  diamonds), where the hard-sphere limit leaves the packings 
  hyperstatic, with coordination numbers that depend on
  $\mu$ and $\epsilon$.}
\label{zvsk}
\end{center}
\end{figure}

We next investigate the effect of the finite stiffness of the spheres
on the packings, and on the average coordination number $z$ in particular.
Let us assume that the packings formed by these stiff elastic spheres are 
statistically equivalent to packings that would be obtained by first forming 
a truly hard-sphere packing and subsequently allowing elastic relaxation. 
Due to the slight compression of the spheres of finite stiffness under gravity,
we expect to see an increase in coordination number during this elastic 
relaxation. Since the typical compressive strain of a sphere under the same 
loading conditions scales as $1/k_n$, we expect that a finite fraction of 
neighbors in the hard-sphere packing that were within a distance $d[1+O(mg/k_n d)]$ 
of each other form new contacts upon elastic relaxation. The number of
such near-contacts in the hard-sphere packing can be computed by integrating
the hard-sphere RDF $g_\infty(r)$ over the range $r/d \in (1,1+mg/k_n d)$, 
yielding an effective coordination number
\begin{equation}
z_n(k_n) =z_{\infty} +{\tilde a}_n \left(\frac{k_n d}{mg}\right)^{-\alpha_n},
\ \frac{k_n d}{mg} \gg 1,
\label{eqzn}
\end{equation}
where $z_\infty$ is the coordination number in the hard-sphere limit,
${\tilde a}_n$ is a constant, and the exponent [cf. Eq.(\ref{eqgrlimit})]
\begin{equation}
\alpha_n=1-\alpha.
\label{expiden}
\end{equation}
Thus, there is a power-law correction to the 
apparent coordination number, with an exponent that depends on the nature 
of the singularity in the first peak of $g(r)$.  A numerical fit of the 
data to Eq. (\ref{eqzn}) shown in Fig.~\ref{zvsk}(a) results in 
$\alpha_n=0.498\pm0.002$ and $z_\infty=6.01\pm0.02$, in excellent
agreement with the isostaticity hypothesis, as well as the exponent 
relation Eq. (\ref{expiden}). Makse {\it et al.} also found 
that $z$ approaches 6 as the stress goes to zero in their numerical 
studies of compressed spheres\cite{makse1}. 

Armed with this insight, we apply a similar analysis to frictional
packings, with results presented in Fig.~\ref{zvsk}(b). Although 
the RDF of frictional packings appears to have the same square-root 
divergence near $r=d$ (see Fig.~\ref{grfirstpeak}), the numerical fit 
to Eq. (\ref{eqzn}) in the presence of friction yields a different
exponent $\alpha_f\approx -1/4$, resulting in a slower approach to the 
hard-sphere limit. Thus, the exponent identity Eq. (\ref{expiden}) does not
hold for frictional spheres. Moreover, in contrast to the frictionless case, 
the hard-sphere limits remain firmly above the isostatic value of four,
and vary as a function of $\mu$ and $\epsilon$.

Even though the rather unlikely scenario of a further crossover to 
isostaticity at extreme stiffnesses cannot be entirely ruled out, 
it must be pointed out that the stiffest spheres with
$k_{n}=2\cdot10^{9}mg/d$ in these packings experience strains
$\delta/d \alt 10^{-8}$. This should be compared to the strain of a
``typical'' grain, i.e. a glass sphere with a 100 micron diameter,
under just its own weight on the Earth's surface: $\delta/d\approx(\rho
g d/E)^{2/3}$.  With the Young's Modulus $E\approx 6\cdot10^{10}$ Pa,
$\rho\approx 2\cdot10^3 {\rm~kg/m}^3$ and $g\approx 10$ m/s$^2$, this
strain is about $10^{-7}$. Thus, even if isostaticity is ultimately
restored, the relevance of the hard-sphere limit for real granular
systems is still questionable.

\subsection{Plasticity of Contacts}

\label{secplastic}

\begin{figure}[h]
\begin{center}
\includegraphics[width=7.5cm]{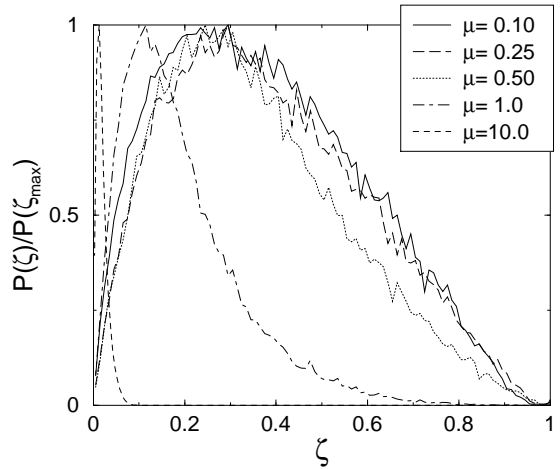}
\caption{Probability distribution $P(\zeta)$, normalized by its 
maximum value, for various values of $\mu$. Curves for $\mu\geq1$ would 
collapse if plotted against $F_t/F_n$ instead of $\zeta$.}
\label{zetaplot}
\end{center}
\end{figure}

One potential explanation for the hyperstaticity of these frictional 
packings is the loss of degrees of freedom for the tangential forces 
in contacts that have become ``plastic'' such that $F_{t}=\mu F_{n}$. 
If a finite fraction of the contacts satisfied this condition, the 
isostaticity condition would need to be modified\cite{leo10}.  However, as
demonstrated by the distribution of the plasticity index $\zeta\equiv
\frac{F_{t}}{\mu F_{n}}$ of the contacts in Fig.~\ref{zetaplot},
almost all contacts in the static packings are below their frictional
threshold $\zeta=1$, eliminating this possibility. A similar
distribution of $\zeta$ was observed for a static packing created from
flow arrest \cite{leo10}. 

For $\mu\geq1$, the distribution of the contact force ratio 
$F_t/F_n$ indeed becomes independent of $\mu$, which manifests itself
as a collapse of $P(\zeta)$ when plotted against $F_t/F_n$ (not shown). 
This result is in accord with the observation in Sec.~\ref{seccoord} that 
packings in this range of $\mu$ behave effectively the same as for systems 
with $\mu=\infty$.

\subsection{Effect of dissipation and initial conditions}

The dependence of the packing geometry on coefficient of restitution
$\epsilon$ is also interesting. Unlike $k_n$ and $\mu$, changing
$\epsilon$ would not change the configuration of a static packing
after it has stopped -- it only affects the relaxation dynamics 
by increasing the removal rate of kinetic energy. 
In this sense, changing $\epsilon$ is like changing the
quench rate of a supercooled liquid as it undergoes a glass
transition. For very large quench rates, the system might be expected
to stop immediately upon forming the minimum number of contacts
necessary to achieve static mechanical equilibrium. 

\begin{figure}[h]
\begin{center}
  \includegraphics[width=7.5cm]{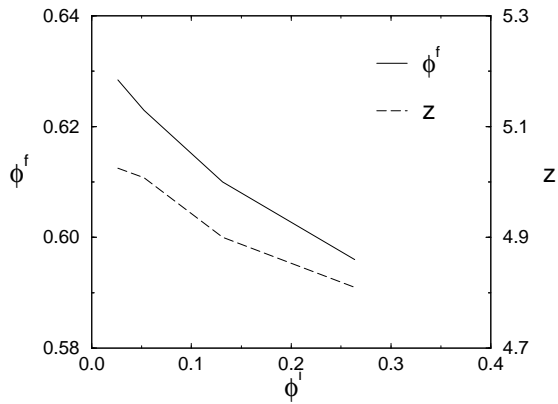} 
\caption{Dependence of the final packing fraction $\phi^{f}$ and coordination
  number $z$ on the initial packing fraction of the falling particles
  $\phi^{i}$, for $k_{n}=2\cdot10^{5}mg/d,~\epsilon=0.88$, and $\mu=0.50$.}
\label{figure7}
\end{center}
\end{figure}

Similarly to the effect of $\epsilon$, we find that the initial
starting densities also affect the final packing. In Fig.~\ref{figure7} we
plot the variation in the final packing fraction $\phi^{f}$, and
coordination number $z$, as a function of the starting density
$\phi^{i}$. We find that more dilute starting states lead to more
compact final states. This behaviour may be due to the increase in
potential energy the system receives when it is more dilute,
converting into kinetic energy of the particles during settling, and
enabling them to explore more of the phase space on their way to
achieving a preferred packing. An empirical fit to the packing
fraction is given by,
\begin{equation}
\phi^{f}=0.5778+0.0567\exp(-4.3\phi^{i}),
\end{equation}
which is similar to the empirical fit for model 2D systems proposed in
Ref.~\cite{blumenfeld1}. However, it should be noted that for 
frictional spheres, extrapolation of the coordination number to the 
limit $\phi^i=\phi^f$ does not result 
in isostaticity as observed in Ref.~\cite{blumenfeld1}.

In light of the dependence of the final state on such parameters as $\epsilon$
and $\phi^i$, the breakdown of the exponent identity Eq.(\ref{expiden})
for frictional packings is perhaps not surprising. Packings obtained by
the sedimentation of hard-spheres followed by elastic relaxation are probably
not statistically equivalent to packings of particles that are elastic from the 
start, due to the strong history-dependence of the final states obtained.  

In general, for hyperstatic packings the force network is not uniquely 
determined
by the packing and the loadings on the particles. It follows that the 
determination of the force network for hyperstatic packings of perfectly 
rigid particles is an ill-posed problem\cite{deniz1}. Thus the order of
the limit $k_n \to \infty$ and the preparation of the packing cannot be 
commuted for frictional spheres. More bluntly, perfectly rigid sphere systems
with friction are unlikely to capture the mechanical properties of packings of 
frictional, elastic spheres, even in the limit of extremely large rigidity of these 
latter particles.

\section{conclusions}

We have studied large scale packings of spherical
grains of varying hardness, friction coefficient, and coefficient of
restitution, formed by sedimentation. We accounted for the systematic
variation with particle stiffness and were able to infer properties of
hard-sphere packings. Although frictionless hard-spheres
appear to form isostatic packings regardless of construction history
and restitution coefficient, frictional packings achieve a multitude
of hyperstatic packings that depend on system parameters and
construction history \cite{footnote10}. The coordination number reduces
smoothly from $z=6$ as the friction coefficient is increased, contrary
to the hypothesis of isostaticity in such packings.

\acknowledgments

Sandia is a multiprogram laboratory operated by Sandia Corporation, a
Lockheed Martin Company, for the United States Department of Energy
under Contract DE-AC04-94AL85000.  DL acknowledges support from
US-Israel Binational Science Foundation Grant 1999235.


\vspace{-0.2in}

\begin{references}

\bibitem{angell1}
C.~A. Angell, K.~L. Ngai, G.~B. McKenna, P.~F. McMillan, and S.~W. Martin,  J.
  App. Phys. {\bf 88,} 3113 (2000).

\bibitem{weitz1}
J. Liu, D.~A. Weitz, and B.~J. Ackerson,  Phys. Rev. E {\bf 48,} 1106 (1993).

\bibitem{lacasse1}
T.~G. Mason, M.-D. Lacasse, G.~S. Grest, D. Levine, J. Bibette, and D.~A.
  Weitz,  Phys. Rev. E {\bf 56,} 3150 (1997).

\bibitem{alexander1}
S. Alexander,  Phys. Rep. {\bf 296,} 65 (1998).

\bibitem{edwards2}
S.~F. Edwards,  Physica A {\bf 249,} 226 (1998).

\bibitem{grinev1}
S.~F. Edwards and D.~V. Grinev,  Physica A {\bf 263,} 545 (1999).

\bibitem{makse1}
H.~A. Makse, D.~L. Johnson, and L.~M. Schwartz,  Phys. Rev. Lett. {\bf 84,}
  4160 (2000).

\bibitem{leo10}
L.~E. Silbert, D. Erta{\c s}, G.~S. Grest, T.~C. Halsey, and D. Levine,
  cond-mat/0109124 (unpublished).

\bibitem{ohern1}
C.~S. O'Hern, S.~A. Langer, A.~J. Liu, and S.~R. Nagel, cond-mat/0110644.

\bibitem{footnote7}
We use Hookean springs as opposed to the more realistic Hertzian force law
  since for Hookean springs the coefficient of restitution $\epsilon$ is
  constant, whereas it is velocity-dependent for Hertzian spheres.
  Consequently, it is far easier to drain residual kinetic energy from Hookean
  systems en route to a static packing. We have also studied some systems with
  Hertzian contacts. We found that our main conclusion that the coordination
  number reduces smoothly from six with increasing friction holds.

\bibitem{walton1}
O.~R. Walton and R.~L. Braun,  J. Rheo. {\bf 30,} 949 (1986).

\bibitem{cundall1}
P.~A. Cundall and O.~D.~L. Strack,  G$\acute{e}$otechnique {\bf 29,} 47 (1979).

\bibitem{leo7}
L.~E. Silbert, D. Erta{\c s}, G.~S. Grest, T.~C. Halsey, D. Levine, and S.~J.
  Plimpton,  Phys. Rev. E {\bf 64,} 051302 (2001).

\bibitem{wolf1}
J. Schafer, S. Dippel, and D.~E. Wolf,  J. Phys. I France {\bf 6,} 5 (1996).

\bibitem{berryman}
J.~G. Berryman,  Phys. Rev. A {\bf 27,} 1053 (1983).

\bibitem{torquato}
S. Torquato, private communication. The abundance of such near-contacts is
  puzzling. A more detailed investigation of the origin of this singularity,
  using a common-contact analysis akin to the common neighbor analysis
  described in Ref.~\protect\cite{clarke}, but with a cutoff distance of
  $r_c=1^+$, is beyond the scope of this study.

\bibitem{blumenfeld1}
R. Blumenfeld, S.~F. Edwards, and R.~C. Ball, cond-mat/0105348.

\bibitem{deniz1}
T.~C. Halsey and D. Erta{\c s},  Phys. Rev. Lett. {\bf 83,} 5007 (1999).

\bibitem{footnote10}
We also find similar results using Hertzian contacts.

\bibitem{clarke}
A.~S. Clarke and H. Jonsson,  Phys. Rev. E {\bf 47,} 3975 (1993).

\end{references}
\end{document}